\documentclass[preprint]{revtex4}

\usepackage{graphicx}

\begin{document}

\title{Power of a Turbulent Pulsar}

\author{Andrei Gruzinov}
 
\affiliation{Center for Cosmology and Particle Physics, Department of Physics, New York University, NY 10003}

\date{February 25, 2005}

\begin{abstract}

  We speculate that stationary force-free pulsar magnetospheres are screw unstable, and the spin-down power of real pulsars is carried by turbulent Poynting jets. The turbulent jet entrains poloidal flux as it propagates away from the star. Due to poloidal flux entrainment, poloidal field inside the jet decreases at a slower rate than the dipole's $r^{-3}$, and the pulsar power increases accordingly: $L\sim (\mu ^2\Omega^4/c^3)(\Omega R/c)^{-\alpha }$, where $\mu$ is the magnetic dipole, $\Omega$ is the frequency, $R$ is the neutron star radius, and $\alpha \lesssim 1$ is an index which depends on spin-dipole angle. Our speculation is of interest because it seems to provide the only possible explanation (without fine-tuning) of observed pulsar braking indices. 

\end{abstract}

\maketitle

\section{Introduction}
Pulsars spin down. The spin-down power is probably carried by Poynting jets propagating along open magnetic field lines, and the resulting spin-down power is thought to be  
\begin{equation}\label{conv_power}
L=f(\theta) {\mu ^2\Omega^4\over c^3}
\end{equation}
where $f(\theta )\sim 1$ is an unknown dimensionless coefficient depending on the spin-dipole angle $\theta$  \cite{gold}. The power is comparable to the magnetic dipole power \cite{gunn} because one assumes that pulsars simply eject the characteristic dipole field $B_{lc} \sim \mu /r_{lc}^3$ at the light cylinder radius $r_{lc}\equiv c/\Omega$, giving the power $L\sim r_{lc}^2cB_{lc}^2\sim \mu ^2\Omega^4/c^3$. 

The conventional pulsar power formula (\ref{conv_power}) has been confirmed recently, when the axisymmetric pulsar magnetosphere was calculated, giving the dimensionless coefficient for aligned rotators \cite{ cont, gruz}: $f(0)=1\pm 0.1$. Assuming that non-zero $\theta$ solutions are similar to the aligned case, it would seem that the large-scale electromagnetic background, on which all the pulsar emission is occurring, has basically been established.

One, however, can argue differently. Stationary pulsar magnetospheres can be unstable. Then real pulsar magnetospheres are turbulent, and to account for the effects of turbulence, one has to modify the pulsar power formula. Within the framework of FFE (force-free electrodynamics, see eg. \cite{gruz} and references therein), the only dimensionless parameter, which one can use to modify the pulsar power, is $\Omega R/c<<1$, where $R\sim 10$km is the neutron star radius. Since the background is self-similar, the only reasonable modification of the pulsar power is the power law. Thus, we will postulate that the correct pulsar power is given by
\begin{equation}\label{power}
L=f{\mu ^2\Omega^4\over c^3}\left({\Omega R\over c}\right) ^{-\alpha},
\end{equation}  
where the dimensionless coefficient $f\sim 1$ and index $\alpha \lesssim 1$ depend on the spin-dipole angle $\theta$. The negative power in equation (\ref{power}) corresponds to poloidal field entrainment into the Poynting jet. This leads to an increased value of the field within the jet, and thus to a larger power.

If one uses the anomalous pulsar power formula (\ref{power}) to infer the surface magnetic fields of neutron stars, the predicted values of the fields will be lower than the standard ones. Also, the turbulent magnetosphere model might be needed to understand the pulsar emission. The most important reason to take the turbulent model seriously is because it provides the only viable explanation of pulsar timing observations, \S2.

\section{Pulsar timing and anomalous pulsar power}
Pulsar spin-down may be quantified by braking indices
\begin{equation}\label{indices}
n\equiv {\Omega \ddot{\Omega }\over \dot {\Omega } ^2},~~~~~m\equiv {\Omega ^2 \dddot{\Omega }\over \dot {\Omega } ^3}.
\end{equation}  
The values of the first braking index $n$ are less than 3 in all five pulsars where it has been measured \cite{livi}. Even more important for our purposes is the fact that for Crab ($n=2.519$, $m=10.23\pm 0.03$ from Table 1 of \cite{lyne}) the second braking index is $m=n(2n-1)$ to $2\sigma$. Unfortunately, this result can possibly be ``contaminated by timing noise and frequent glitches'' \cite{livi}. 

What are the theoretical expectations for the braking indices \cite{blan} ? 

Anomalous pulsar power (\ref{power}), with the assumption of no change in the spin-dipole angle and no change in the dipole moment, gives the spin-down equation 
\begin{equation}\label{spin-down}
\dot{\Omega }\propto - {\Omega }^{3-\alpha }
\end{equation}  
in agreement with both $3>n=3-\alpha$ and $m=n(2n-1)$.

Conventional pulsar power (\ref{conv_power}) gives $n=3$ for constant spin-dipole angle and dipole moment. One needs to assume that the angle or/and the dipole moment change. Let us consider these possibilities.

The model postulating variable dipole moment requires fine-tuning, because there is no reason to expect that the time scale for the change of the magnetic field and the time scale for the spin-down should be the same. If one accepts Crab's $m=n(2n-1)$, the fine-tuning problem becomes much more severe.

The model postulating evolution of the spin-dipole angle might seem to require no fine-tuning. It is known \cite{davi} that magnetic dipole rotating in vacuum aligns in about the spin-down time. However, real stars are not spherical tops, and real magnetic fields are not ideal dipoles. Due to magnetic multipoles in the torque formula (and also magnetic and elastic deformations of the ellipsoid of inertia of the star \cite{gold1}), the time scale for spin-dipole angle evolution  is expected to be shorter than the spin-down time by a factor $\sim \Omega R/c \ll 1$ (or even smaller factor coming from the ellipsoid of inertia). Thus, this theory cannot work without fine-tuning. With Crab's $m=n(2n-1)$, the fine-tuning problem becomes much more severe.

We are aware of only one model which requires no fine-tuning in explaining $n\neq 3$. The model postulates that during spin-down, the outward-moving  superfluid vortex tubes entrain the superconducting magnetic flux tubes \cite{saul}. This automatically gives the right time scale for the magnetic field evolution. However: (i) this theory still requires fine-tuning if one accepts Crab's $m=n(2n-1)$, (ii) this theory allows both $n<3$ and $n>3$, depending on the initial field configuration.

Anomalous pulsar power appears to be the most natural explanation of pulsar timing observations. To confirm or refute this model, one needs to perform a time-dependent three-dimensional FFE simulation of a pulsar magnetosphere. The analytical methods only say that screw is the most promising instability \cite{gruz1}. But even the problem of FFE stability of an axisymmetric pulsar, for which the precise shape of the stationary magnetosphere is known, looks too difficult for analytical methods (the problem is not self-adjoint; there are singular current layers; the equatorial current layer is not an FFE current layer).

\begin{acknowledgments}
I thank Yuri Levin and Mal Ruderman for useful discussions. This work was supported by the David and Lucile Packard Foundation.
\end{acknowledgments}

\end{document}